\documentclass[aip, pof,graphicx]{revtex4-2}

\usepackage{graphicx}% Include figure files
\usepackage{dcolumn}% Align table columns on decimal point
\usepackage{bm}% bold math
\usepackage[utf8]{inputenc}
\usepackage[T1]{fontenc}
\usepackage{amssymb}
\usepackage{mathptmx}
\usepackage{color}
\usepackage{longtable}
\usepackage{epstopdf}%This line makes .eps figures into .pdf - please comment out if not required.

\draft % marks overfull lines with a black rule on the right

\begin{document}

%\preprint{}

\title[Crystallization and refluidization in very-narrow fluidized beds]{Crystallization and refluidization in very-narrow fluidized beds\\
	This article may be downloaded for personal use only. Any other use requires prior permission of the author and AIP Publishing. This article appeared in Phys. Fluids 35, 093306 (2023) and may be found at https://doi.org/10.1063/5.0163555.} %Title of paper



\author{Vin\'icius P. S. Oliveira}
\affiliation{School of Mechanical Engineering, UNICAMP - University of Campinas,\\
	Rua Mendeleyev, 200, Campinas, SP, Brazil
}

\author{Danilo S. Borges}
\affiliation{School of Mechanical Engineering, UNICAMP - University of Campinas,\\
	Rua Mendeleyev, 200, Campinas, SP, Brazil
}

\author{Erick M. Franklin}%
 \email{erick.franklin@unicamp.br}
 \thanks{Corresponding author}
\affiliation{School of Mechanical Engineering, UNICAMP - University of Campinas,\\
	Rua Mendeleyev, 200, Campinas, SP, Brazil
}


\date{\today}

\begin{abstract}
Fluidization of solid particles by an ascending fluid is frequent in industry because of the high rates of mass and heat transfers achieved. However, in some cases blockages occur and hinder the correct functioning of the fluidized bed. In this paper, we investigate the crystallization (defluidization) and refluidization that take place in very-narrow solid-liquid fluidized beds under steady flow conditions. For that, we carried out experiments where either monodisperse or bidisperse beds were immersed in water flows whose velocities were above those necessary for fluidization, and the ratio between the tube and grain diameters was smaller than 6. For monodisperse beds consisting of regular spheres, we observed that crystallization and refluidization alternate successively along time, which we quantify in terms of macroscopic structures and agitation of individual grains. We found the characteristic times for crystallization, and propose a new macroscopic parameter quantifying the degree of bed agitation. The bidisperse beds consisted of less-regular spheres placed on the bottom of a layer of regular spheres (the latter was identical to the monodisperse beds tested). We measured the changes that macroscopic structures and agitation of grains undergo, and show that the higher agitation in the bottom layer hinders crystallization of the top layer. Our results bring new insights into the dynamics of very-narrow beds, in addition to proposing a way of mitigating defluidization.
\end{abstract}

\pacs{}% insert suggested PACS numbers in braces on next line

\maketitle %\maketitle must follow title, authors, abstract and \pacs

\section{INTRODUCTION}
\label{sec:intro}

The use of fluidized beds, in which solid particles are fluidized by an ascending fluid, is frequent in industry because of the high rates of mass and heat transfers achieved. In general, the solid is fragmented into small particles (grains), placed in a vertical tube, and a fluid is forced to flow upwards with velocities adjusted to maintain the granular bed suspended. However, in some cases blockages can occur and hinder the correct functioning of the fluidized bed.

One of the main difficulties for fully understanding fluidized beds is the presence of different scales in a given bed, with clusters, plugs, crystals and other structures being found at the bed scale, and solid-solid contacts occurring at the grain scale. Other difficulty is the absence of scale invariance. For large or normal beds, those for which the ratio of tube $D$ to grain $d$ diameter is $D/d$ $\gtrsim$ 100, structures such as bubbles and slugs usually appear \cite{Geldart,Sundaresan,guazzelli_book,Koralkar}. For narrow beds, those for which 10 $<$ $D/d$ $\lesssim$ 100, transverse waves, blobs and bubbles appear \cite{Duru,Duru2}. For very-narrow beds, those for which $D/d$ $\lesssim$ 10,  plugs, crystallization (in the sense of defluidization) and jamming can occur \cite{Cunez, Cunez2, Cunez3, Cunez4}. Another problem is that the bed dynamics depends on the fluid state (gas or liquid): while solid-solid collisions and fluid drag are the most pertinent mechanisms \cite{koch,Sundaresan} in gas-solid fluidized beds, virtual mass and pressure forces can be as important in solid-liquid fluidized beds (SLFBs) \cite{Cunez,Cunez2}.

The dynamics and granular structures in large gas-solid beds were extensively investigated over the last decades \cite{Miller, Geldart, Rietema, Menon, Escudero}, but fewer works investigated narrow SLFBs \cite{Anderson, ElKaissy, Didwania, Zenit, Zenit2, Duru, Duru2, Goldman, Aguilar, Ghatage}. In general terms, the latter showed that confinement effects change the granular structures to transverse waves and blobs. In particular, Goldman and Swinney \cite{Goldman} investigated narrow SLFBs undergoing defluidization. For that, they started with a fluidized bed and decreased the water flow until reaching velocities still above that for minimum fluidization ($U_{mf}$). Afterward, they increased slightly the water velocity. They showed a two-stage process: (i) glass formation, (called here crystallization or defluidization), corresponding to a static structure with small fluctuations (at the level of grains) that appears during the deceleration phase (i.e., no macroscopic motion); and (ii) jamming, corresponding to a jammed structure without small fluctuations (no microscopic motion) that appears after increasing slightly the fluid velocity once the bed is crystallized. In addition, they showed that bed crystallization depends on the deceleration rate.

There are still fewer investigations on very-narrow SLFBs. C\'u\~nez and Franklin \cite{Cunez,Cunez2} carried out experiments and numerical simulations using CFD-DEM (computational fluid dynamics - discrete element method) for beds with $D/d$ $<$ 6, and observed  alternating high- and low-compactness regions, called plugs and bubbles, in the bed. They showed that a dense network of contact forces exists within the granular plugs, percolating from their centerline to the tube wall, which evinces a dynamics under the effects of high confinement. In addition, C\'u\~nez and Franklin \cite{Cunez2} investigated particle segregation and layer inversion in beds consisting of two species, showing how confinement affetcs the inversion. Latter, C\'u\~nez and Franklin \cite{Cunez3} investigated experimentally very-narrow SLFBs under partial defluidization and showed that crystallization can happen when fluid velocities are above $U_{mf}$. However, different from the narrow bed of Ref. \cite{Goldman}, they found that distinct granular lattices appear depending on the grain type, that crystallization does not depend on the deceleration rate, and that the jamming intensity depends on the grain type. More recently, C\'u\~nez et al. \cite{Cunez4} investigated experimentally and numerically very-narrow SLFBs consisting of bonded spheres (duos and trios), and showed that the motion of particles and the bed structures are different from those found with loose grains. 

By carrying out resolved CFD-DEM computations, Yao et al. \cite{Yao} investigated numerically monodisperse and bidisperse beds in a duct with square cross section with $L/d_L$ $=$ 6, where $L$ is the square length and $d_L$ the diameter of the largest grains. Since the domain was periodic in the transverse coordinates, it does not correspond to a narrow case. The authors found that most of properties (packing fraction, particle fluctuations, kinematic wave speed, and collisional and hydrodynamic stresses) of segregated layers in the bidisperse cases are roughly the same as in the corresponding monodisperse cases. However, they showed that granular fluctuations in the upper layer are the largest at low Reynolds numbers, while fluctuations in the transition and lower layers are the largest at moderate and high Reynolds numbers, respectively. Although not strictly a narrow case, some of their results may still be valid for highly confined beds, but this remains to be investigated.

The use of SLFBs is also frequent in biological reactors that involve mass transfers \cite{Dempsey,Nelson}, which are narrow or very-narrow depending on the dimensions of the used tube and on the growth of an organic film around each particle. As a matter-of-fact, crystallization and jamming cannot occur for the correct functioning of those reactors. In this paper, we inquire further into the crystallization (defluidization) and refluidization that occur in very-narrow SLFBs under steady flow conditions. For that, we carried out experiments where either monodisperse or bidisperse beds were  immersed in water flows whose velocities were above those necessary for fluidization, and we filmed the bed and processed the images, obtaining measurements at both the bed and grain scales. For monodisperse beds consisting of regular spheres, we observe that crystallization and refluidization alternate successively along time, which we quantify in terms of duration, macroscopic structure, and degree of agitation. We found the characteristic time for crystallization, and propose a new macroscopic parameter quantifying the degree of bed agitation. The bidisperse beds consisted of less-regular spheres placed on the bottom of a layer of regular spheres (that were the same used in the monodisperse beds), and we did not observe crystallization. We measured the macroscopic structures and the degree of agitation, and show that the higher agitation of the bottom layer hinders crystallization in the top layer. Other than bringing new insights into the dynamics of very-narrow beds, our results show an effective way of mitigating defluidization issues observed in narrow and very-narrow SLFBs. 

In the following, Sec. \ref{sec:exp_setup} presents the experimental setup, Sec.  \ref{sec:results} shows the results and Sec. \ref{sec:conclusions} concludes the paper.

\section{EXPERIMENTAL SETUP}
\label{sec:exp_setup}

\begin{figure}[ht]
	\centering
	\includegraphics[width=0.7\columnwidth]{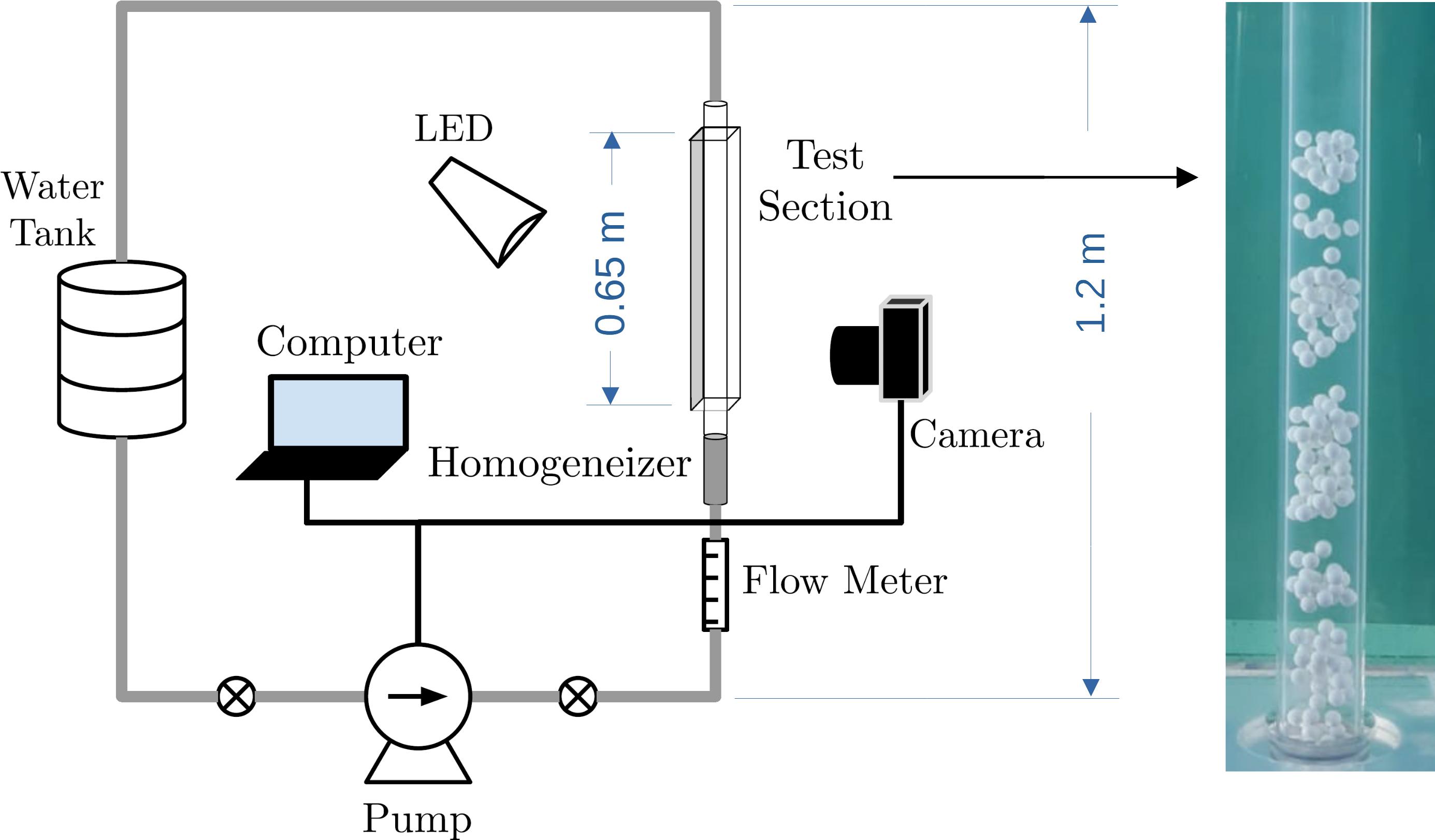}
	\caption{Layout of the experimental setup.}
	\label{fig:1}
\end{figure} 

The experimental setup consisted basically of a water tank, a centrifugal pump, a flow meter, a flow homogenizer, a 25.4-mm-ID vertical tube, and a return line. The water flowed in closed loop (in the order just described), with the flow rate controlled by a frequency inverter connected to a computer. The vertical tube was transparent (made of polymethyl methacrylate - PMMA), 1.2 m long, and aligned vertically within $\pm 3^{\circ}$. Its first 0.65 m (downstream the homogenizer) corresponded to the test section, and a visual box filled with water was placed around it to minimize optical distortions. The flow homogenizer consisted of a 150-mm-long tube filled with $d$ = 6 mm spheres packed between wire screens. Figure \ref{fig:1} shows the layout of the experimental setup and a photograph of the test section.

\begin{figure}[ht]
	\centering
	\includegraphics[width=0.5\columnwidth]{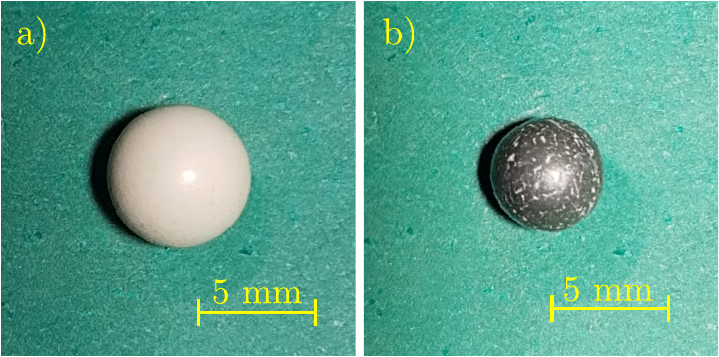}
	\caption{(a) Polymer covered spheres and (b) aluminum spheres used in the experiments.}
	\label{fig:particles}
\end{figure}

For the liquid, we used tap water within 25$^{\circ}$C $\pm$ 3$^{\circ}$C, so that its density $\rho_f$ and dynamic viscosity $\mu_f$ were approximately 1000 kg/m$^3$ and 10$^{-3}$ Pa.s, respectively. For the grains, we used 0.2 g polymer-covered spheres ($\rho_{p1}$ = 1768 kg/m$^3$) with diameter $d_1$ = 5.95 mm $\pm$ 0.01 mm, and aluminum spheres ($\rho_{p2}$ = 2760 kg/m$^3$) with diameter $d_2$ = 4.8 mm $\pm$ 0.03 mm, which we call species 1 and 2, respectively (photographs are shown in Fig. \ref{fig:particles}). In this way, $D/d_1$ = 4.3 and $D/d_2$ = 5.3, and the numbers of Stokes $St_t \,=\, v_t d \rho_p / (9\mu_f)$ and Reynolds $Re_t \,=\, \rho_f  v_t d / \mu_f$ based on the terminal velocity were, respectively, 404 and 2056 for species 1 and 696 and 2269 for species 2, where $v_t$ is the terminal velocity of one single sphere. The values of $St_t$ and $Re_t$ (much higher than unity) indicate that the solid particles (grains) have much more inertia than fluid particles of same volume.

Prior to each test, grains were let to settle in the test section, forming a granular bed. Two different beds were investigated: a monodisperse bed with $N_1$ = 200 particles of species one, and a bidisperse bed consisting of a top layer with $N_1$ = 200 particles of species 1 and a bottom layer with $N_2$ = 300 particles of species 2. Therefore, the top layer of the bidisperse bed has the same composition of the monodisperse bed. With the grains settled, water flows with cross-sectional average velocities $U$ within 0.06 and 0.12 /ms were imposed by controlling the rotation of the centrifugal pump. For each bed type, Tab. \ref{table:table1} presents the average particle fraction $\phi_0$ of the settled bed, the heights $H_{if}$ and water velocities $U_{if}$ at the inception of fluidization, and the settling velocity of spheres $v_{s,i}$ (for each species). Values of $v_{s,i}$ were computed with the Richardson--Zaki correlation, $v_{s,i} = v_t \left( 1-\phi_0 \right) ^{2.4}$, and those of $\phi_0$, $H_{if}$ and $U_{if}$ were determined experimentally by using image processing, where $\phi_0$ is given by $H_{if}$ multiplied by the tube cross section and divided by the total volume of particles (volume occupied by the solid phase). The total duration of tests was of either 300 or 600 s.

\begin{table}[ht]
	\caption{Type of bed, number of particles $N_1$ and $N_2$, particle fraction $\phi_0$ of the settled bed, bed height $H_{if}$ and water velocity $U_{if}$ at the inception of fluidization, and settling velocity of spheres of species 1 and 2, $v_{s,i,1}$ and $v_{s,i,2}$.}
	\label{table:table1}
	\centering
	\begin{tabular}{c c c c c c c c}  
		\hline\hline
		Type & $N1$ & $N2$ & $H_{if}$ & $\phi_0$  & $v_{s,i,1}$ & $v_{s,i,2}$ & $U_{if}$\\
		$\cdots$& $\cdots$ & $\cdots$ & m & $\cdots$ & m/s & m/s & m/s\\ [0.5ex] % inserts table 
		%heading 
		\hline % inserts single horizontal line 
		Monodisperse & 200 & 0 & 0.082  & 0.531  & 0.093  & $\cdots$ & 0.045 \\
		Bidisperse & 200 & 300 & 0.082  & 0.531  & 0.093  & 0.091 & 0.045 \\
		\hline
		\hline 
	\end{tabular}
\end{table}

Since we had optical access to the tube interior, we made use of digital images in our investigation. For that, we placed a camera of complementary metal-oxide-semiconductor (CMOS) type perpendicularly to the test section (lateral view of the bed). The camera had a resolution of 1920 px $\times$ 1080 px when operating at 60 Hz, and it was branched to a computer controlling both the camera and the pump rotation. We mounted a lens of $60$ mm focal distance and F2.8 maximum aperture on the camera, and set the camera frequency to 60 Hz and the region of interest (ROI) to 725 px $\times$ 165 px, for a field of view of 111.6 mm $\times$ 25.4 mm. With that, 1 px corresponds to approximately 0.16 mm in our images. To have stable illumination while avoiding undesirable reflections from the tube, we placed either a black or a green background, used lamps of light-emitting diode (LED) branched to a continuous-current source, and placed translucent papers just in front of LED lamps. Once acquired and stored, the images were processed with numerical codes written in the course of this work.

\subsection{Image Processing}

\begin{figure}[ht]
	\centering
	\includegraphics[width=0.6\columnwidth]{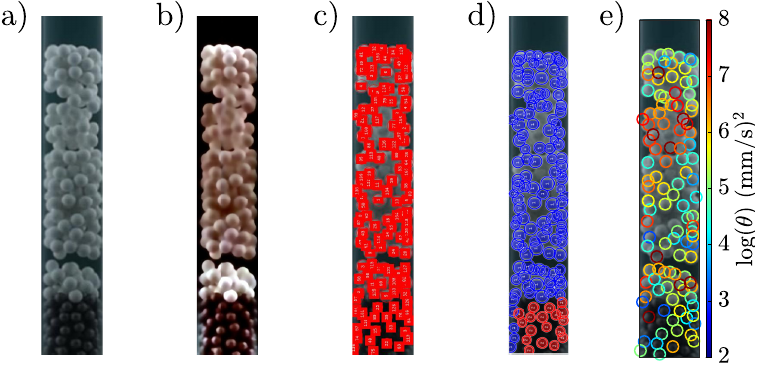}
	\caption{Some steps of the applied image processing: (a) raw image; (b) filter enhancement; (c) particle detection; (d) particle discrimination; (e) granular temperature.}
	\label{fig:scheme_image_processing}
\end{figure}

The acquired movies were processed for obtaining measurements at both the bed and grain scales. For that, we first segmented the movies into individual frames. Then, on each frame, we selected the ROI, carried out a background construction, re-scaled the image size, removed the background, and applied bilateral filtering and contrast enhancement. Afterward, we identified the granular bed and bed structures (plugs) by using the horizontally averaged intensity of pixels (a value of 50\% of intensity was considered in the transition between plugs and bubbles). Next, we identified individual grains (and their positions) by applying the Hough transformation \cite{Yuen} and discriminated each species based on the pixel intensity. Finally, we tracked each grain along frames by using the Hungarian algorithm \cite{Munkres} (for the time and space correlation) and the Kalman filter \cite{Kalman} (for noise attenuation in positions and velocities). Some steps of the image processing applied in this work can be seen in Fig. \ref{fig:scheme_image_processing}.

Since we only have access to grains in contact with the tube wall, our measurements are based on a plane projection of a cylindrical surface, which we associate with a Cartesian coordinate system. With that procedure, the instantaneous velocities of grains were decomposed into their transverse $x$ and longitudinal $y$ components, $V_x$ and $V_y$, respectively. We then computed ensemble averages of the $x$ and $y$ components of velocity, $V_{x,avg}$ and $V_{y,avg}$, respectively, where we consided all particles (in the bed or of a given species, depending on the case). Afterward, we computed the corresponding fluctuations, $v'_x$ = $V_x - V_{x,avg}$ and $v_y$ = $V_y - V_{y,avg}$, respectively, for each image pair. With these fluctuations, we computed the granular temperature of each particle, 

\begin{equation}
	\theta_p = \frac{1}{2}\left( (v'_x)^2 + (v'_y)^2 \right) \,\,.
	\label{eq_temp}
\end{equation}

Next, we estimated the horizontal averages of the granular temperature, $\theta(t)$, by dividing the bed into vertical regions wherein we computed the ensemble average of the granular temperature (taking into account only grains in contact with the tube wall, for optical access reasons). Although differences are expected, we associate the horizontal averages with cross-sectional averages in the tube.

\section{RESULTS}
\label{sec:results}

By processing the images, we were able to identify macroscopic structures and quantities, such as the bed height, interface between layers, celerities of surfaces, granular plugs, and lattices (see Fig. \ref{fig:snapshots_heights} for examples of macroscopic measurements -- Multimedia available online), and microscopic quantities as well, such as the motion of individual particles (mean flow and fluctuation). In general terms, we observed differences between monodisperse beds and the top layer of bidisperse beds, which we present next.

\subsection{Monodisperse beds}

As in Refs. \cite{Cunez, Cunez3}, we observed the formation of granular plugs that moved upwards and, after some time had elapsed, that the bed crystallized under some flow velocities. However, different from Ref. \cite{Cunez3}, some of the monodisperse beds (i) always crystallized and (ii) refluidized and crystallized successively over time. These differences are perhaps due to the different spheres used (polymer covered) and longer duration of the present tests when compared with those of Ref. \cite{Cunez3}.

\begin{figure}[ht]
	\centering
	\includegraphics[width=0.70\linewidth]{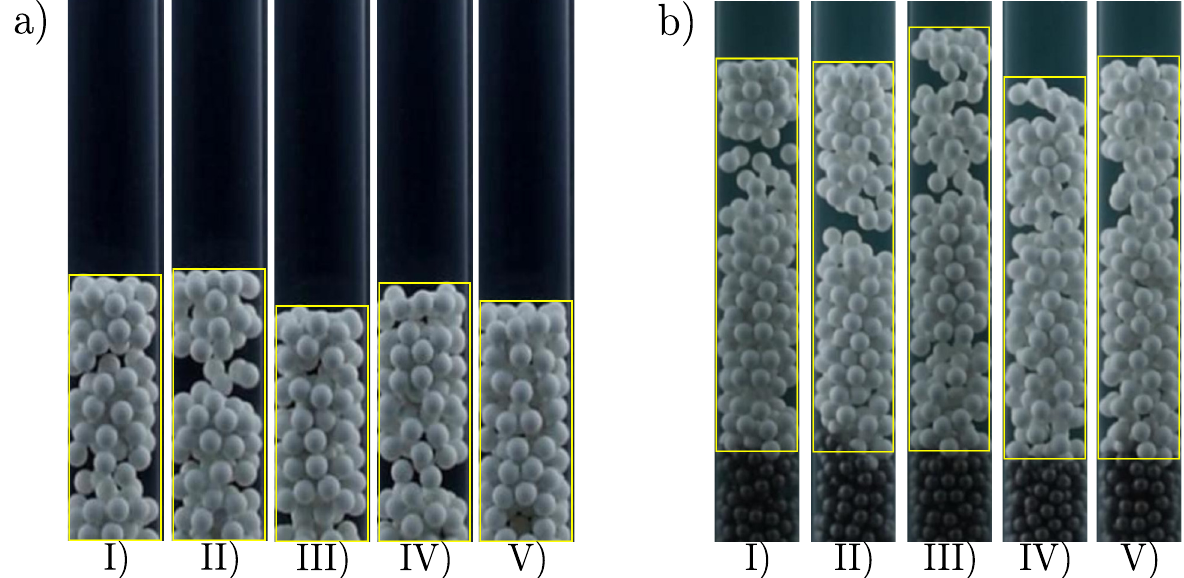}\\
	\caption{Snapshots showing the evolution of the height of (a) a monodisperse bed (case p of Tab. \ref{tab:summary}) and (b) the top layer of a bidisperse bed (case e$_2$ of Tab. \ref{tab:summary}). The interval between frames is of 0.5 s. Multimedia available online}
	\label{fig:snapshots_heights}
\end{figure}

\begin{figure}[ht]
	\centering
	\includegraphics[width=0.99\columnwidth]{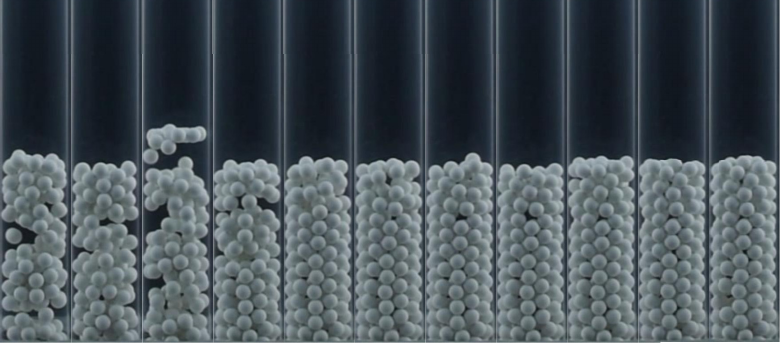}
	\caption{Snapshots of a monodisperse bed, showing the process of crystallization (case q of Tab. \ref{tab:summary}). The time between frames is 4.5 s. Multimedia available online.}
	\label{fig:snapshot_crystallization}
\end{figure}

In terms of macroscopic quantities, we processed the images to measure the time evolution of the bed height, plug length, plug celerity, and packing fraction $\phi$. We then computed time averages of the height, plug length and plug celerity, which we call $H$, $\lambda$ and $C$, respectively. The average is suitable for the analysis since $H$, $\lambda$ and $C$ oscillate around a mean value due to the passage of plugs when the bed is fluidized. To identify if the bed is either fluidized or crystallized, we computed the rate of variation of the mean packing fraction $\phi_{float}$,

\begin{equation}
	\phi_{float} = \left\{ max\left(\frac{d \phi_{y}}{dt}\right)-\overline{\frac{d\phi_{y}}{dt}}\right\} \,\,,
	\label{eq_phi_float}
\end{equation} 

\noindent where $\phi_y$ is the horizontally averaged packing fraction (function of $t$ and $y$) and $\overline{\frac{d\phi_y}{dt}}$ is the time derivative of $\phi_y$ averaged in the vertical direction. $\phi_{float}$ represents thus the upper bound of time variations of the mean packing fraction (packing-fraction rate), being higher for fluidized beds in comparison with crystallized beds. This permits the identification of crystallization, such as that shown in the snapshots of Fig. \ref{fig:snapshot_crystallization} (a crystallized bed is also visible in the last frame of Fig. \ref{fig:snapshots_heights}a -- Multimedia available online).

\begin{figure}[ht]
	\centering
	\includegraphics[width=0.8\columnwidth]{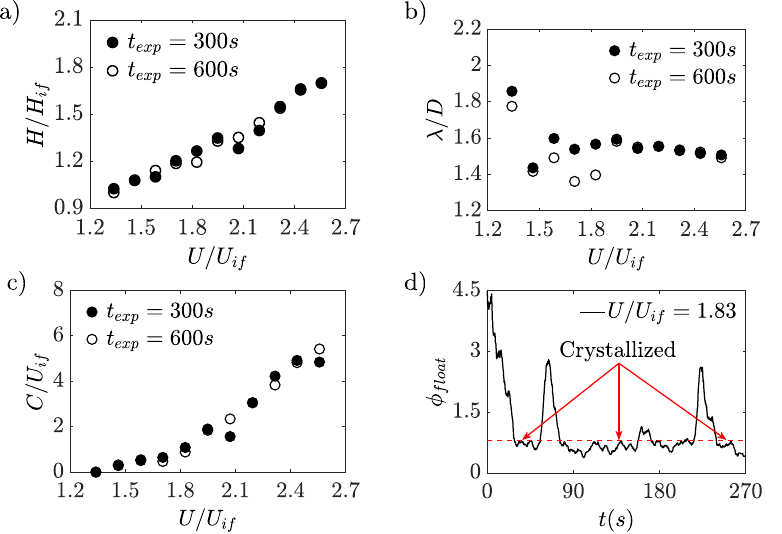}
	\caption{Characteristics of monodisperse beds. (a) Dimensionless height $H/H_{if}$; (b) plug length $\lambda / D$; and (c) plug celerity $C/U_{if}$ as functions of the dimensionless cross-sectional average velocity $U/U_{if}$. (d) Time evolution of the packing-fraction rate $\phi_{float}$. Panel (d) corresponds to case p and panels (a)-(c) to cases a-v of Tab. \ref{tab:summary}, and the red-dashed line corresponds to $\phi_{float}$ = 0.8.}
	\label{fig:graph_mono}
\end{figure}

Figures \ref{fig:graph_mono}a--\ref{fig:graph_mono}d show, respectively, how the height $H$, plug length $\lambda$ and plug celerity $C$ vary with the cross-sectional average velocity $U$, and an example of time evolution of the packing-fraction rate $\phi_{float}$. In Figure \ref{fig:graph_mono}, $H$, $C$ and $U$ are normalized by the respective values at incipient fluidization ($H_{if}$ and $U_{if}$), and $\lambda$ is normalized by the tube diameter $D$. Beginning with the bed height $H$, Fig. \ref{fig:graph_mono}a shows a tendency of growth, though slightly deviations are observed when $ U/U_{if}$ is within 1.8 and 2.4. For the plug length $\lambda$ and celerity $C$, Figs. \ref{fig:graph_mono}b and \ref{fig:graph_mono}c show that $\lambda$ is roughly independent of $U$ (although a considerable scatter is observed in Fig. \ref{fig:graph_mono}b), while the celerity increases with $U$, in agreement with the results of Ref. \cite{Cunez}. Finally, Fig. \ref{fig:graph_mono}d shows the time evolution of $\phi_{float}$ for a monodisperse bed with $U$ = 0.0822 m/s (case p of Tab. \ref{tab:summary}), from which we observe that the bed alternates between fluidized and crystallized states, crystallization corresponding to $\phi_{float}$ $\approx$ 0.8 (graphics for other cases are available in the supplementary material). This alternation between fluidized and crystallized states has never been reported, and we investigate it next in terms of characteristic times and degree of agitation. In addition, we inquire into a method to mitigate crystallization in Subsection \ref{Bidisperse beds}.

\begin{figure}[ht]
	\centering
	\includegraphics[width=0.4\columnwidth]{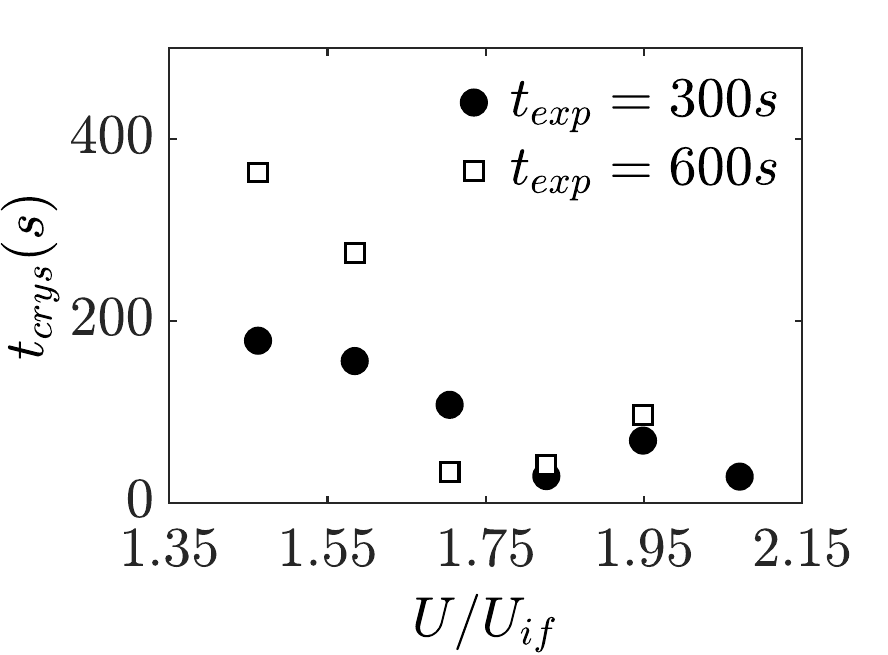}
	\caption{Crystallization time $t_{crys}$ as a function of $U/U_{if}$ for the tests with $t_{exp}$ = 300 and 600 s (cases a-v of Tab. \ref{tab:summary}).}
	\label{fig:graph_tc}
\end{figure}

Based on $\phi_{float}$, it is possible to obtain the crystallization time $t_{crys}$, which is defined as the typical time that a bed, once fluidized, takes until reaching crystallization. For that, we computed ensemble averages of the time intervals corresponding to $\phi_{float}$ $>$ 0.8 in the $\phi_{float}$ datasets. The results are summarized in Fig. \ref{fig:graph_tc}, showing $t_{crys}$ as a function of $U/U_{if}$ for the 300 and 600 s experiments. We observe that $t_{crys}$ initially decreases sharply with the flow velocity, and from $U/U_{if}$ = 1.75 on it increases slightly or remains roughly constant. These observations can be drawn despite a significant deviation of values between the 300 and 600 s experiments for $U/U_{if}$ $<$ 1.75, which occur due to the lower number of samples in the 300 s experiments. The non-monotonical behavior (or, at least, the asymptotic behavior toward a constant value) is not \textit{a priori} expected: the increase in agitation with increasing flow rates should decrease the probabilities of crystallization. Therefore, within the ranges of water velocities in which crystallization occurs, the increase in the flow rate is not an effective means of reducing crystallization. 

A summary of the macroscopic results can be found in Tab. \ref{tab:summary}, which lists, for each case, the cross-sectional average velocity $U/U_{if}$, bed height $H/H_{if}$, standard deviation of the bed height $\sigma_H/H_{if}$, plug length $\lambda/D$, standard deviation of the plug length $\sigma_{\lambda}/D$, and celerity $C/U_{if}$ in dimensionless forms, as well as the crystallization time $t_c$ and duration of the test run $t_{exp}$. A table with a summary of data in dimensional form is available in the supplementary material.

\setlength{\tabcolsep}{6pt}

\begin{longtable}{ccccccccccc}
	\caption{Summary of macroscopic measurements. For each case, the table shows the type of bed (M = monodisperse and B = bidisperse), dimensionless cross-sectional average velocity $U/U_{if}$, dimensionless bed height considering species 1 only (top layer in case of bidisperse beds) $H/H_{if}$, dimensionless standard deviation of the bed height $\sigma_H/H_{if}$, dimensionless plug length $\lambda/D$, dimensionless standard deviation of the plug length $\sigma_{\lambda}/D$, dimensionless celerity $C/U_{if}$, crystallization time $t_c$, and duration of the test run $t_{exp}$.}
	\label{tab:summary} \\
	\hline \hline
	\centering
	Case & Type & $U/U_{if}$ & $H/H_{if}$ & $\sigma_H/H_{if}$ & $\lambda/D$ & $\sigma_{\lambda}/D$ & $C/U_{if}$ & $\sigma_C/U_{if}$ & $t_c$ (s) & $t_{exp}$ (s)\\
	\hline
	
	\endfirsthead
	
	\hline
	\multicolumn{11}{c}{Table \ref{tab:summary} continued}\\
	\hline
	Case & Type & $U/U_{if}$ & $H/H_{if}$ & $\sigma_H/H_{if}$ & $\lambda/D$ & $\sigma_{\lambda}/D$ & $C/U_{if}$ & $\sigma_C/U_{if}$ & $t_c$ (s) & $t_{exp}$ (s)\\
	\hline
	\endhead
	
	\hline \hline
	\endfoot
	
	a & M & 1.34 & 1.00 & 0.0181 & 1.78 & 0.0565 & 0.000 & 0.000 & - & 600 \\ 
	b & M & 1.46 & 1.08 & 0.0182 & 1.42 & 0.1786 & 0.289 & 0.361 & 364 & 600 \\ 
	c & M & 1.58 & 1.14 & 0.0267 & 1.49 & 0.2893 & 0.510 & 0.988 & 275 & 600 \\ 
	d & M & 1.71 & 1.19 & 0.0207 & 1.36 & 0.2414 & 0.472 & 0.462 & 34 & 600 \\ 
	e & M & 1.83 & 1.19 & 0.0208 & 1.40 & 0.2295 & 0.883 & 3.100 & 42 & 600 \\ 
	f & M & 1.95 & 1.33 & 0.0459 & 1.58 & 0.4002 & 1.840 & 5.709 & 97 & 600 \\ 
	g & M & 2.07 & 1.35 & 0.0533 & 1.55 & 0.4043 & 2.343 & 7.213 & - & 600 \\ 
	h & M & 2.19 & 1.45 & 0.0663 & 1.56 & 0.4358 & 3.055 & 8.943 & - & 600 \\ 
	i & M & 2.31 & 1.54 & 0.0708 & 1.53 & 0.4336 & 3.836 & 10.816 & - & 600 \\ 
	j & M & 2.44 & 1.66 & 0.0818 & 1.53 & 0.4199 & 4.821 & 13.240 & - & 600 \\ 
	k & M & 2.56 & 1.70 & 0.0842 & 1.49 & 0.4290 & 5.421 & 14.708 & - & 600 \\ 
	l & M & 1.34 & 1.03 & 0.0159 & 1.86 & 0.0492 & 0.002 & 0.236 & - & 300 \\ 
	m & M & 1.46 & 1.08 & 0.0255 & 1.44 & 0.300 & 0.318 & 0.513 & 179 & 300 \\ 
	n & M & 1.58 & 1.10 & 0.0349 & 1.60 & 0.3013 & 0.544 & 0.889 & 156 & 300 \\ 
	o & M & 1.71 & 1.20 & 0.0344 & 1.54 & 0.3774 & 0.648 & 2.126 & 108 & 300 \\ 
	p & M & 1.83 & 1.27 & 0.0856 & 1.57 & 0.3789 & 1.085 & 2.477 & 30 & 300 \\ 
	q & M & 1.95 & 1.35 & 0.0471 & 1.60 & 0.3801 & 1.898 & 6.505 & 69 & 300 \\ 
	r & M & 2.07 & 1.28 & 0.0874 & 1.54 & 0.3452 & 1.566 & 3.598 & 29 & 300 \\ 
	s & M & 2.19 & 1.40 & 0.1799 & 1.56 & 0.4358 & 3.055 & 8.943 & - & 300 \\ 
	t & M & 2.31 & 1.55 & 0.0703 & 1.53 & 0.4261 & 4.227 & 11.806 & - & 300 \\ 
	u & M & 2.44 & 1.66 & 0.0831 & 1.52 & 0.4441 & 4.918 & 13.512 & - & 300 \\ 
	v & M & 2.56 & 1.70 & 0.0846 & 1.51 & 0.3810 & 4.844 & 13.521 & - & 300 \\ 
	a$_2$ & B & 1.34 & 0.99 & 0.1867 & 2.99 & 0.4582 & 0.783 & 7.069 & - & 300 \\ 
	b$_2$ & B & 1.46 & 1.14 & 0.0274 & 2.41 & 1.6510 & 3.945 & 10.147 & - & 300 \\ 
	c$_2$ & B & 1.58 & 1.17 & 0.0366 & 2.95 & 1.8097 & 8.467 & 29.850 & - & 300 \\ 
	d$_2$ & B & 1.71 & 1.15 & 0.0472 & 2.37 & 1.8100 & 11.069 & 34.173 & - & 300 \\ 
	e$_2$ & B & 1.83 & 1.37 & 0.0614 & 2.37 & 0.9625 & 6.612 & 19.463 & - & 300 \\ 
	f$_2$ & B & 1.95 & 1.26 & 0.0660 & 1.88 & 0.6794 & 5.672 & 15.817 & - & 300 \\ 
	g$_2$ & B & 2.07 & 1.40 & 0.0876 & 1.72 & 0.5803 & 5.887 & 17.138 & - & 300 \\ 
	h$_2$ & B & 2.19 & 1.54 & 0.1005 & 1.79 & 0.6597 & 5.968 & 18.665 & - & 300 \\ 
	i$_2$ & B & 2.31 & 1.61 & 0.1034 & 1.64 & 0.5331 & 6.053 & 18.043 & - & 300 \\ 
	j$_2$ & B & 2.44 & 1.72 & 0.1286 & 1.63 & 0.5500 & 6.109 & 19.403 & - & 300 \\ 
	k$_2$ & B & 2.56 & 1.82 & 0.1283 & 1.56 & 0.4956 & 6.636 & 19.888 & - & 300 \\
	
\end{longtable}

\begin{figure}[ht]
	\includegraphics[width=0.8\columnwidth]{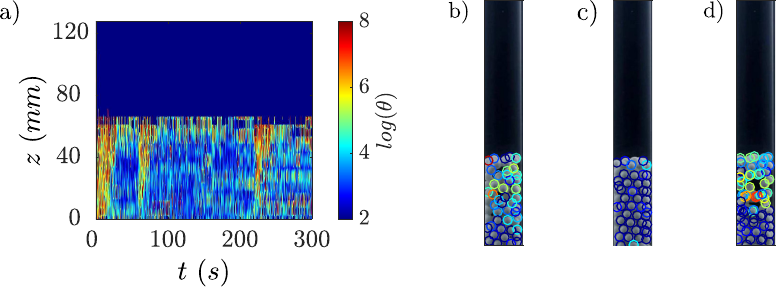}
	\caption{(a) Spatiotemporal diagram of granular temperature $\theta$ for case p. (b)--(d) Snapshots showing grains with their corresponding values of $\theta_p$ for $t$ = 63, 125 and 225 s. The colorbar on the right of panel (a) shows values of 10$\log \theta$ (instead of $\theta$) to accentuate differences.}
	\label{fig:gran_temp_mono}
\end{figure}

Figure \ref{fig:gran_temp_mono}a shows a spatio-temporal diagram of cross-sectional averages of the granular temperature $\theta$ for case p, and Figs. \ref{fig:gran_temp_mono}b-d present some snapshots showing grains with their corresponding values of $\theta_p$. Figures \ref{fig:gran_temp_mono}b--d show that particles have some degree of fluctuation in $t$ = 63 s (Fig. \ref{fig:gran_temp_mono}b) and 225 s (Fig. \ref{fig:gran_temp_mono}d), that in $t$ = 125 s (Fig. \ref{fig:gran_temp_mono}c) particles have very low values of $\theta_p$ (crystallized bed), and that crystallization initiates at the bottom of the bed (visible in the bottom of Figs. \ref{fig:gran_temp_mono}b and \ref{fig:gran_temp_mono}d). This tendency of starting crystallization from below (in agreement with Ref. \cite{Cunez3}) is also noticeable in Fig. \ref{fig:gran_temp_mono}a, which also shows the alternating crystallization and refluidization of the bed.

\begin{figure}[ht]
	\includegraphics[width=0.8\columnwidth]{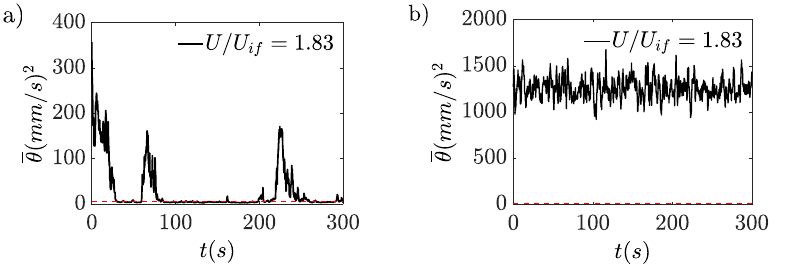}
	\caption{Ensemble-averaged granular temperature $\overline{\theta}$ for (a) a monodisperse (case p of Tab. \ref{tab:summary}) and (b) the top layer of a bidisperse bed (case e$_2$ of of Tab. \ref{tab:summary}). Crystallized and fluidized regions can be compared directly with Figs. \ref{fig:graph_mono}d and \ref{fig:graph_bi}d. The red-dashed line corresponds to a reference value when the bed is crystallized ($\overline{\theta}$ = 5 (mm/s)$^2$).}
	\label{fig:avg_gran_temp_case}
\end{figure}

\begin{figure}[ht]
	\includegraphics[width=0.8\columnwidth]{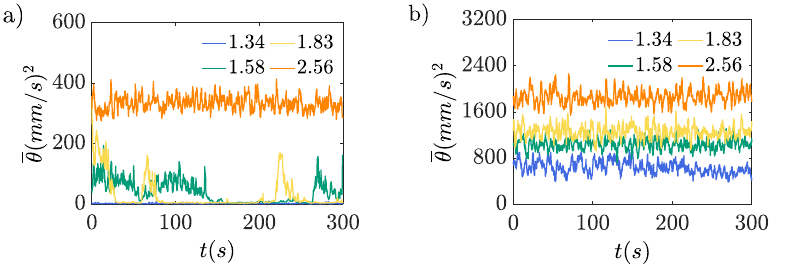}
	\caption{Ensemble-averaged granular temperature $\overline{\theta}$ for (a) monodisperse beds, cases l, n, p and r; and (b) top layer of bidisperse beds, cases a$_2$, c$_2$, e$_2$ and k$_2$. Captions correspond to values of $U/U_{if}$.}
	\label{fig:avg_gran_temp}
\end{figure}

To inquire further into the level of bed agitation, we computed, at each time instant, an ensemble average of the granular temperature of the bed, $\overline{\theta}$,

\begin{equation}
	\overline{\theta}_j = \frac{1}{N_j} \sum_{1}^{N_j} \theta_j \,\,,
\end{equation}

\noindent where $N_j$ is the number of particles of species $j$. Figure \ref{fig:avg_gran_temp_case}a shows the time evolution of $\overline{\theta}$ for case p, where we observe a behavior in agreement with that of $\phi_{float}$, low values ($\overline{\theta}$ $\approx$ 5 (mm/s)$^2$) corresponding to crystallized states and higher values to fluidized states. Therefore, $\phi_{float}$ indicates whether the bed is fluidized or crystallized without the necessity of measurements at the grain scale. Although it seems straightforward, this is an important result since $\phi_{float}$ is a macroscopic parameter, easily measurable even with low quality cameras and/or transducers.

Finally, we compare the time evolution $\overline{\theta}$ for different water velocities $U$, as shown in Fig. \ref{fig:avg_gran_temp}. We can directly observe the existence of a range for fluidization and crystallization, since for $U$ $\leq$  0.060 m/s the bed is settled, and for $U$ $\geq$ 0.099 m/s the bed is always fluidized (crystallization does not occur). We note that we used polymer-coated particles in our experiments, and that, perhaps, the alternation between crystallization and fluidization (and crystallization times) can be different for other types of particles. The effect of the particle surface on crystallization-refluidization problems remains to be investigated.

\subsection{Bidisperse beds}
\label{Bidisperse beds}

We investigate now the behavior of the previous beds (species 1) when placed as a layer on the top of another one made up of grains of a different type (species 2). Grains of species 2 do not crystallize under the water velocities used in the monodisperse tests, so that the basic idea is to inquire if the agitation of the bottom layer hinders crystallization in the top layer.

\begin{figure}[ht]
	\centering
	\includegraphics[width=0.8\columnwidth]{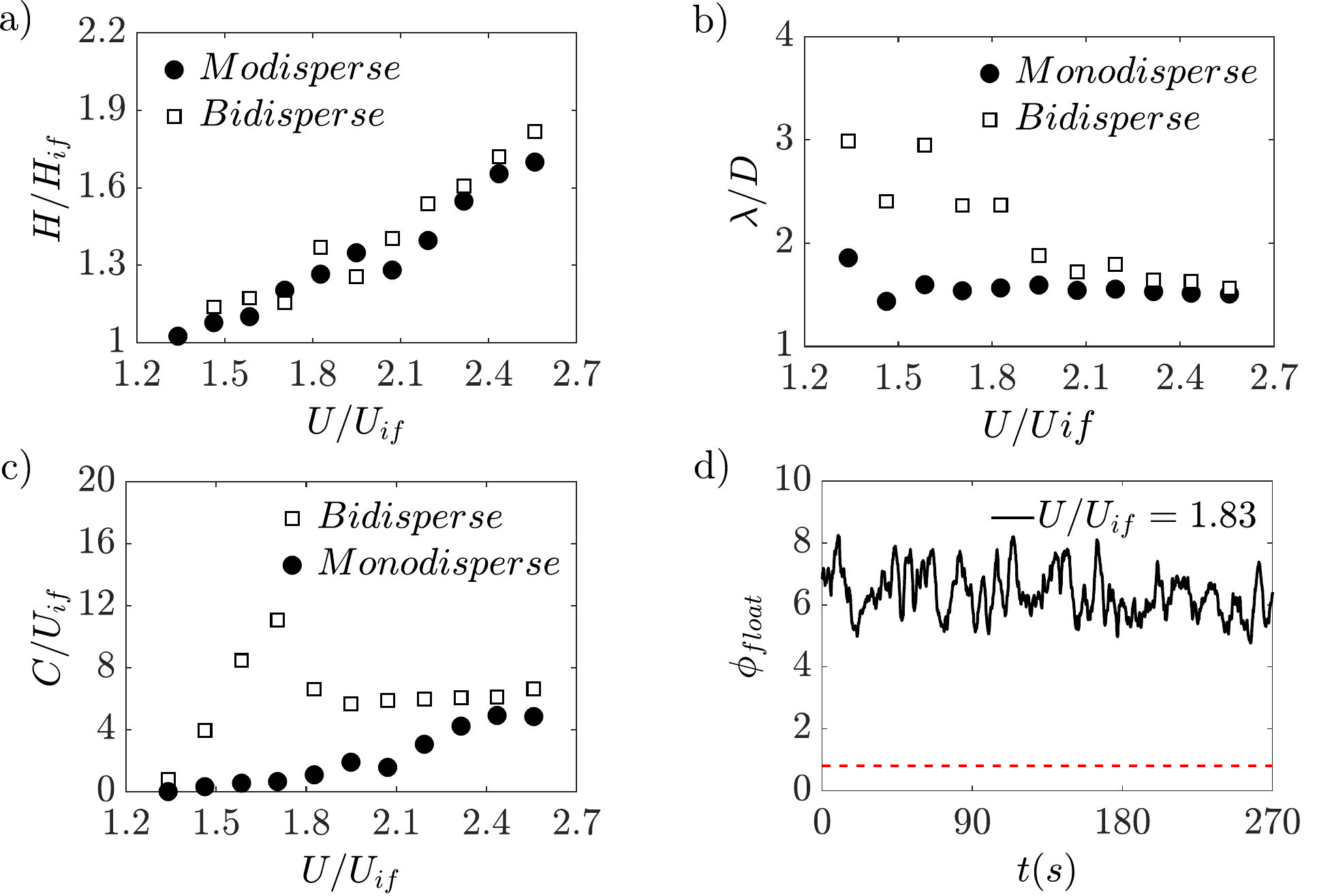}
	\caption{Characteristics of bidisperse beds. (a) Dimensionless height of species 1 (top layer in case of bidisperse bed) $H/H_{if}$; (b) plug length of species 1 $\lambda / D$ ; and (c) plug celerity $C/U_{if}$ of species 1 as functions of the cross-sectional average velocity $U/U_{if}$. (d) Time evolution of the packing-fraction rate $\phi_{float}$. In panels (a)-(c) values corresponding to monodisperse beds are shown for comparison. Panel (d) corresponds to case e$_2$ of Tab. \ref{tab:summary}, and the red-dashed line shows $\phi_{float}$ = 0.8.}
	\label{fig:graph_bi}
\end{figure}

Figures \ref{fig:graph_bi}a-c show the height $H$, plug length $\lambda$ and plug celerity $C$ for species 1 only (top layer in case of bidisperse beds) as functions of the cross-sectional average velocity $U$ for both bidisperse (open squares) and monodisperse (solid circles) beds (monodisperse beds are shown for comparison). Although $H/H_{if}$ for bidisperse and monodisperse cases have roughly the same values and trend, corresponding to similar expansions of the bed/layer, the plug length is considerably affected by the bottom layer, in particular for water velocities lower than $U/U_{if}$ $\approx$ 2. While the monodisperse case presents a slight variation for $U/U_{if}$ $\lessapprox$ 1.5, with a constant value of $\lambda / D$ $\approx$ 1.5 for $U/U_{if}$ $\gtrapprox$ 1.5, the bidisperse case shows higher values ($\lambda / D$ $\approx$ 3) for $U/U_{if}$ $\lessapprox$ 1.6, and a decrease with $U/U_{if}$ until reaching what seems an asymptotic value $\lambda / D$ $\approx$ 1.5, identical to that of the monodisperse case, for $U/U_{if}$ $\gtrapprox$ 2.1. For the plug celerities, those of bisdisperse beds present a non-monotonical behavior, increasing and then decreasing with $U/U_{if}$ for $U/U_{if}$ $\lessapprox$ 2, and reaching a value of $C/U_{if}$ $\approx$ $6$ for $U/U_{if}$ $\gtrapprox$ 2, which is roughly the higher celerity reached by the monodisperse bed at the highest water velocity tested. Therefore, in terms of the macrostructures only, the main effect of placing a non-crystallizing layer at the bottom of the crystal-susceptible one is to increase the size and celerity of structures for water velocities lower than  $U/U_{if}$ $\approx$ 2.1. For higher water velocities, the macrostructures tend asymptotically to those observed for monodisperse beds. In their numerical simulations, Yao et al. \cite{Yao} found that most properties of the top layer are approximately equal to those of the corresponding monodisperse case. Our results show that, however, in very-narrow beds this is the case only for $U/U_{if}$ $\gtrapprox$ 2.1.

Figure \ref{fig:graph_bi}d shows an example (case e$_2$) of time evolution of the packing-fration rate $\phi_{float}$ of the top layer, where the red line corresponds to $\phi_{float}$ = 0.8 (which is the typical value of crystallized beds). It is clear from Fig. \ref{fig:graph_bi}d that the mean packing fraction of the top layer remains at much higher values when in the presence of a non-crystallizing layer at the bottom, and that no crystallization has taken place during the test. The same was observed for the other bidisperse cases (graphics for the other cases are available in the supplementary material), indicating that crystallization of a given bed is hindered by placing less regular grains below it.

The reason for hindering crystallization seems, thus, associated with the degree of agitation of the bottom layer. To investigate that, we computed the granular temperature of each particle, $\theta_p$, and the ensemble-average granular temperature $\overline{\theta}$ of the top layer (graphics of the ensemble-average granular temperature of the entire bed are available in the supplementary material). Figure \ref{fig:avg_gran_temp_case}b shows an example of $\overline{\theta}$ for the top layer of a bidisperse bed (case e$_2$), where we observe that throughout the test values of the ensemble-average granular temperature oscillate around $\overline{\theta}$ $\approx$ 1200 (mm/s)$^2$, i.e., much higher than in the monodisperse case. The same behavior was observed for all cases tested. For example, Fig. \ref{fig:avg_gran_temp}b shows $\overline{\theta}$ for the top layer of the bidisperse cases a$_2$, c$_2$, e$_2$ and k$_2$, where we observe that even the lowest water velocity engenders values of $\overline{\theta}$ much higher than in the monodisperse cases.

Therefore, the bottom layer extracts energy from the water flow and transmits part of it to the top layer via the agitation of its grains. This is in agreement with the picture showed by Yao et al. \cite{Yao}, in which fluctuations of grains are higher in the lower and transition (layer-layer interface) regions for moderate and high Reynolds numbers. However, different from previous works, we show that: (i) for a bed susceptible of undergoing crystallization, a bottom layer can hinder crystallization; (ii) the mechanism for avoiding crystallization is the energy transmission via grain-grain interactions, engendered by the agitation of bottom-layer grains. Placing a layer of less regular grains on the bottom of a layer susceptible to crystallization can, thus, mitigate the problem without changing or rebuilding the fluidization facility. Another potential solution could be changing the distributor, but this is out of the scope of this work, in addition of being a more onerous solution.

We stress that, in strict terms, our results show the behavior of particles in contact with the tube wall. The behavior in the core of the bed needs, thus, to be investigated further, which can be done by using RIM (refractive index matching). Besides, resolved CFD-DEM computations can also reveal the dynamics in the core of the bed.

\section{CONCLUSIONS}
\label{sec:conclusions}

In this paper, we inquired into the crystallization and refluidization that occur in very-narrow solid-liquid fluidized beds under steady flow conditions. We basically found that: (i) monodisperse beds consisting of regular spheres crystallize and refluidize successively along time; (ii) when fluidized, the bed consists of granular plugs that propagate upwards; (iii) when crystallized, the bed structure consists of an organized lattice that remains static in the macroscopic scale; (iv) within the tested parameters, the characteristic time for crystallization is between approximately 30 and 360 s; (v) bidisperse beds for which the bottom layer consists of less regular grains do not crystallize; (vi)  the size and celerity of bed structures in the top layer are larger and faster in comparison with those of the corresponding monodisperse case under low water velocities, and tend asymptotically to those of monodisperse beds when under high water velocities; (vii) the mechanism for avoiding crystallization is the energy transmission via the agitation of the bottom-layer grains, which are less regular and not susceptible to crystallization. Finally, we propose a new macroscopic parameter quantifying the degree of grain-scale agitation that can be easily measured even with low quality cameras and/or transducers. Our results represent a significant step toward understanding the dynamics of very-narrow beds and mitigating defluidization problems.

\section*{AUTHOR DECLARATIONS}
\noindent \textbf{Conflict of Interest}

The authors have no conflicts to disclose

\section*{SUPPLEMENTARY MATERIAL}
See the supplementary material for additional figures and an additional table of our results.

\section*{DATA AVAILABILITY}
The data that support the findings of this study are openly available in Mendeley Data at https://doi.org/10.17632/mj5xs54nbh \cite{Supplemental3}.

\begin{acknowledgments}
The authors are grateful to FAPESP (Grant Nos. 2018/14981-7, 2020/00221-0 and 2022/01758-3) and to CNPq (Grant No. 405512/2022-8) for the financial support provided.
\end{acknowledgments}

\bibliography{references}

\end{document}